\newcommand\percc{\ifmmode{\rm cm^{-3}}\else{cm$^{-3}$}\fi}
\newcommand\km{\ifmmode{\rm km}\else{km}\fi}
\newcommand\kms{\km\,\pers}
\newcommand\cmtwo{\ifmmode{\rm cm^{-2}}\else{cm$^{-2}$}\fi}
\newcommand\htwo{\ifmmode{{\rm H}_2}\else{H$_2$}\fi}
\newcommand\pmev{\ifmmode{\rm MeV^{-1}}\else{MeV$^{-1}$}\fi}
\newcommand\ghz{\ifmmode{\rm GHz}\else{GHz}\fi}
\newcommand\pers{\ifmmode{\rm s^{-1}}\else{s$^{-1}$}\fi}
\newcommand\pdeg{\fdg}
\newcommand\lsun{\ifmmode{L_\odot}\else{$L_\odot$}\fi}
\newcommand\gsim{\ga}

\def\ee#1{\ifmmode{\times 10^{#1}}\else{$\times 10^{#1}$}\fi}
\documentclass[12pt,preprint]{aastex}

\begin{document}

\title{Properties of Nearby Starburst Galaxies Based on their Diffuse
Gamma-ray Emission}

\author{Timothy A. D. Paglione$^{1,2,3}$ 
and Ryan D. Abrahams$^{1,2,3}$}
\email{paglione@york.cuny.edu}
\altaffiltext{1}{Department of Earth \& Physical Sciences, York
College, City University of New York, 94-20 Guy R. Brewer Blvd.,
Jamaica, NY 11451, USA}
\altaffiltext{2}{Department of Physics, Graduate Center of the City
University of New York, 365 Fifth Ave., New York, NY 10016, USA}
\altaffiltext{3}{Department of Astrophysics, American Museum of
Natural History, Central Park West at 79th Street, New York, NY 10024, USA}

\begin{abstract}

The physical relationship between the far-infrared and radio fluxes of
star forming galaxies has yet to be definitively determined.  The
favored interpretation, the ``calorimeter model,'' requires that
supernova generated cosmic ray (CR) electrons cool rapidly via
synchrotron radiation.  However, this cooling should steepen their
radio spectra beyond what is observed, and so enhanced ionization
losses at low energies from high gas densities are also required.
Further, evaluating the minimum energy magnetic field strength with
the traditional scaling of the synchrotron flux may underestimate the
true value in massive starbursts if their magnetic energy density is
comparable to the hydrostatic pressure of their disks.  Gamma-ray
spectra of starburst galaxies, combined with radio data, provide a
less ambiguous estimate of these physical properties in starburst
nuclei.  While the radio flux is most sensitive to the magnetic field,
the GeV gamma-ray spectrum normalization depends primarily on gas
density.  To this end, spectra above 100 MeV were constructed for two
nearby starburst galaxies, NGC 253 and M82, using $Fermi$ data.  Their
nuclear radio and far-infrared spectra from the literature are
compared to new models of the steady-state CR distributions expected
from starburst galaxies.  Models with high magnetic fields, favoring
galaxy calorimetry, are overall better fits to the observations.
These solutions also imply relatively high densities and CR ionization
rates, consistent with molecular cloud studies.

\end{abstract}


\section{Introduction}

The successful operation of the $Fermi$ Gamma-ray Space Telescope
provides our first opportunity to infer the potential origins of the
diffuse high energy emission from normal galaxies beyond the Milky 
Way \citep[e. g.,][]{butt}.  Recent detections of diffuse gamma-ray
emission from Local Group, starburst, and Seyfert 2 galaxies
demonstrate the system's sensitivity \citep{lat2, m31.gammas,
n4945.gammas}. With further integration time beyond these detection
studies, we can construct gamma-ray spectra in order to constrain
models of the cosmic ray (CR) interactions that give rise to the
emission in these sources.

For starburst galaxies in particular, the normalization of the
gamma-ray spectrum helps to constrain, among other properties, the
volume-averaged gas density affecting the CR distributions.  With
this and the radio normalization, the strength of the so-called
``calorimeter'' model of CR cooling may be determined for these
sources.  At issue is the discrepancy in massive starbursts between
the minimum energy magnetic field strength, estimated from the radio
flux, and the equipartition field strength, estimated from gas surface
density and assuming hydrostatic balance \citep{thompson}.  Given
their large surface densities and the linearity of the FIR-radio
correlation over decades in flux, \citet{thompson} argues for
relatively high field strengths in starbursts.  These values can be an
order of magnitude above the minimum energy field estimates, and thus
alter significantly our understanding of the energy budgets in
starburst galaxies.

In this work, the gamma-ray, radio, and FIR spectra of the archetypal
starburst galaxies NGC 253 and M82 are quantitatively compared to
self-consistent models of the various energy loss and emission 
mechanisms for CRs in normal galaxies.  A broad parameter space is
explored, including density and magnetic field strength, to constrain
the physical conditions of these starburst nuclei that best agree with
their full spectral energy distributions.

\section{Observations and Results}

The data were obtained using the $Fermi$ Large Area Telescope (LAT)
\citep{lat}.  The LAT is a pair-conversion detector sensitive to
photons with energies between roughly 20 MeV and 300 GeV.  The
publicly available $Fermi$ Science Tools \citep{tools} were used, and 
the threads reviewed on the $Fermi$ Science Center web
site\footnote{\url{http://fermi.gsfc.nasa.gov/ssc/data}} for the
binned likelihood analysis were followed.

%Roughly 2.5 yr of LAT data were downloaded from $10\degr$ regions of
%interest around NGC 253 and M82.  To avoid gamma-rays from the Earth's
%limb, a zenith cut of $105\degr$ was invoked.  The rock angle was
%constrained to be less than $52\degr$ and various other data flags
%were set as suggested by the LAT team.  With the large field of view
%of the LAT, the background must be modeled in order to ensure photons
%only associated with the target source are counted.  The 1FGL catalog
%was used as an initial background, and then $gttsmap$ was used to
%search for unmodeled sources that showed up with high significance.
%Three additional sources were identified within $5\degr$ of NGC 253,
%and six were found near M82.  These new sources were modeled with the
%likelihood routine $gtlike$, and incorporated into the background
%model. To recover the photon flux, exposure maps were created in
%various energy bins using the P6\_V3 instrument response functions.
%The final energy binning was settled upon to maximize the
%signal-to-noise and detection significance, based on the test
%statistic (TS), in each bin.

Almost 3.5 yr of Pass 7 LAT data were downloaded from $10\degr$
regions of interest around NGC 253 and M82.  To avoid gamma-rays from
the Earth's limb, a zenith cut of $100\degr$ was invoked in the data
filtering step.  The rock angle was constrained to be less than
$52\degr$ and various other data flags and filters were set as
suggested by the LAT team.  With the large field of view of the LAT,
the background must be modeled in order to ensure photons only
associated with the target source are counted.  The 2FGL catalog was
used as an initial background, and then $gttsmap$ was used to search
for unmodeled sources that showed up with high significance.  Two
additional sources were identified within $5\degr$ of NGC 253, 
and three were found near M82.  These new sources were incorporated
into the background model with the likelihood routine $gtlike$.
Thirty logarithmically-spaced energy bins per were used for the counts
cube and binned exposure map. The spatial binning for both was set to
$0\pdeg1$ per pixel.  An additional $10\degr$ was added to the
exposure map radius to account for the larger PSF at low energies.
The P7SOURCE\_V6 instrument response functions were used.  The final 
energy binning was settled upon to maximize the signal-to-noise and
detection significance, based on the test statistic (TS), in each
bin.

It should be noted that no source with a significant TS was detected
above 100 MeV at the location of M81.  However, the $0\pdeg61$
separation between M82 and M81 is smaller than or comparable to the
point spread function of the LAT below 1 GeV, so they are currently
difficult to resolve.  For this work, it was assumed that none of the
flux from the position of M82 originates from M81.  The FIR and radio
luminosities of M82 are more than an order of magnitude higher than
those of M81, so the gamma-ray contribution from M81 is expected to be
negligible.

\begin{figure}
\plotone{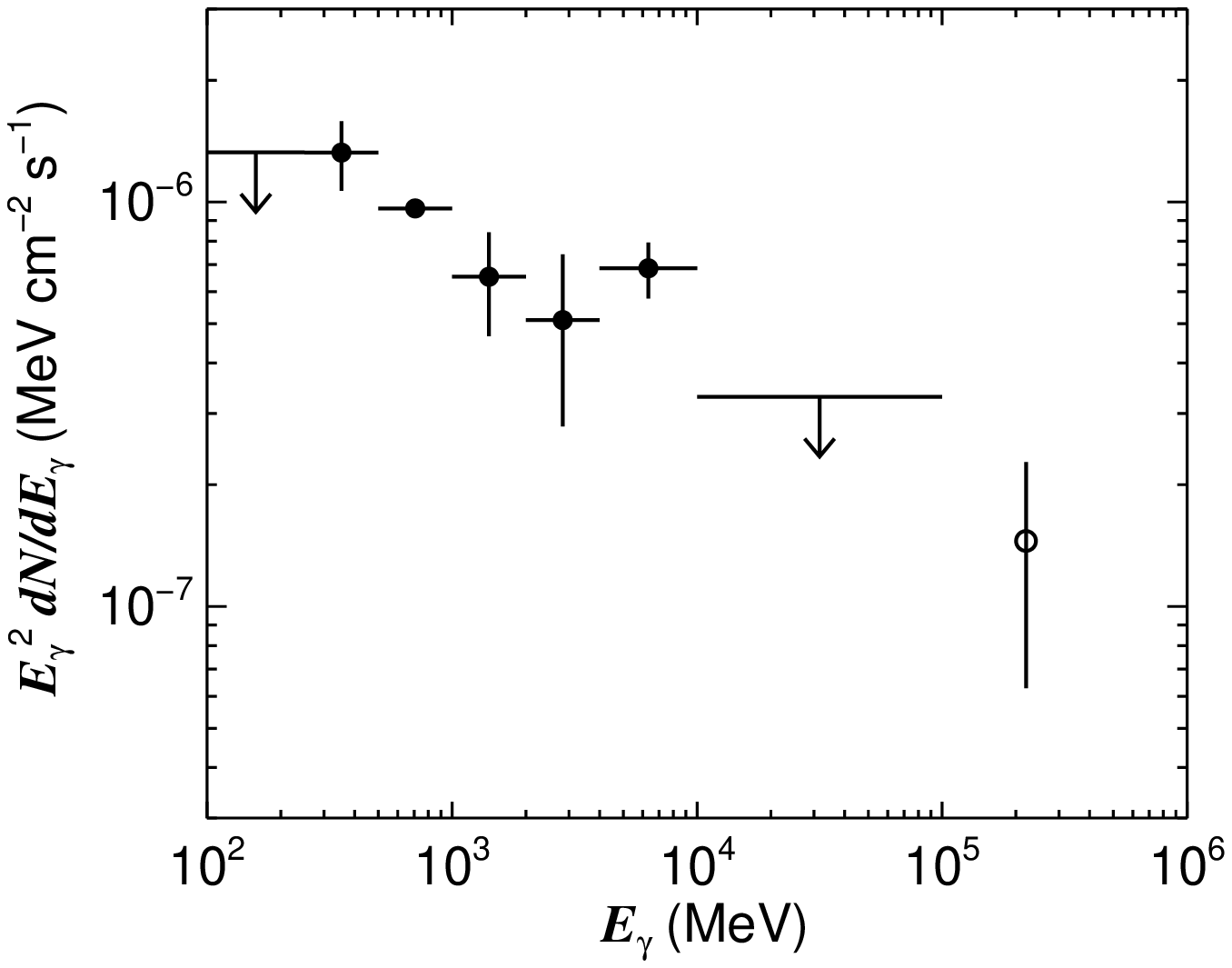}
\caption{Differential gamma-ray spectrum of NGC 253.  Filled symbols
are Fermi data (this work), the open symbol is from HESS
\citep{n253.hess} assuming a photon index of $-2.2\pm 0.3$.
\label{fig.sp253}}
\end{figure}

\begin{figure}
\plotone{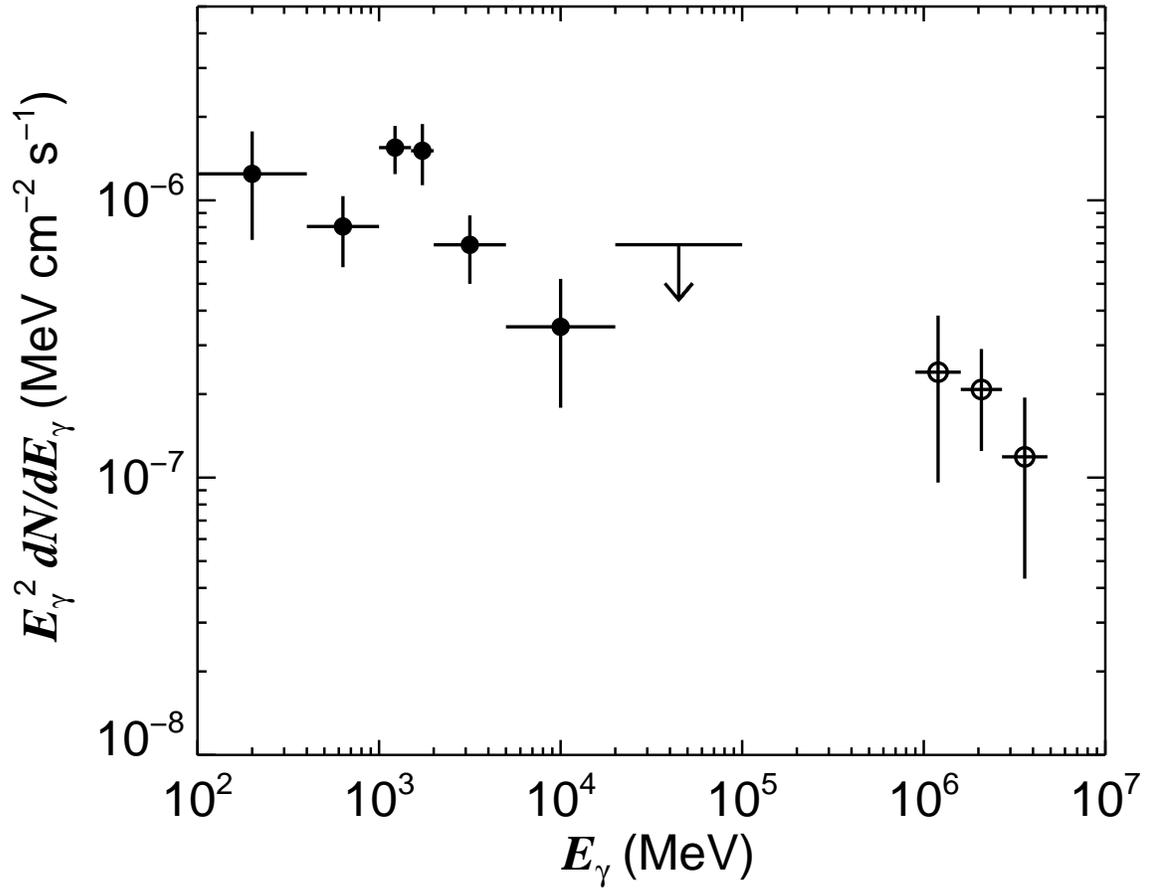}
\caption{Differential gamma-ray spectrum of M82.  Filled symbols
are Fermi data (this work), open symbols are from VERITAS
\citep{m82.veritas}.
\label{fig.sp82}}
\end{figure}

For NGC 253, the integrated flux between 100 MeV and 100 GeV was
$(1.19\pm 0.15)\times 10^{-8}$ ph \cmtwo\ \pers, with a photon 
index of $-2.24\pm0.06$, and TS of 113.  For M82, a total flux of
$(1.44\pm 0.22)\times 10^{-8}$ ph \cmtwo\ \pers\ was found, with a
photon index of $-2.25\pm0.08$, and TS of 180.  These results are
consistent with initial detections \citep{sbg.gammas} and the second
LAT catalog \citep{lat2}.  They also extend to meet the TeV
observations \citep{n253.hess,m82.veritas}. Lightcurves show no
detectable variability in eight-week bins. The differential spectra of
both galaxies are shown in Figures~\ref{fig.sp253} and \ref{fig.sp82},
and are summarized in Table~\ref{tab.spec}.  A linear fit yields
$\chi^2$ values that indicate the spectra are inconsistent with a
simple power law.  In particular, there may be an apparent turnover in
the spectra below a GeV indicative of the expected CR proton
interactions in these sources \citep*{blom}.  The following section
presents models of gamma-ray and radio emission from the expected CR
populations in starburst nuclei.

\section{Modeling the Starburst Emission}\label{sec.sbgmodel}

The likely origin of gamma-rays from starburst galaxies is the
interaction of supernova-accelerated CRs with a large mass of dense,
primarily molecular, gas.  The high temperatures of dense molecular
clouds in starbursts \citep[e. g.,][]{n253.co,sakamoto} indicate
significant penetration and ionization by CRs \citep*{suchkov}.  The
CR electrons themselves emit synchrotron radiation, so radio spectra
of starburst regions serve as a significant constraint as mentioned
before.  The imbedded, dusty star forming regions should radiate
strongly in the FIR, and if the CR electrons lose most of their energy
within them, this would be a natural explanation for the FIR-radio
correlation and calorimetry. However, the gamma-ray spectra are
critical for helping to reduce degeneracies and ambiguities in the
physical solutions.

To quantify the starburst contribution to the gamma-ray and radio
emission from starburst galaxies, the steady-state CR distributions
were modeled.  The models are significantly updated versions of
earlier studies \citep{n253.gammas}, similar in approach to more
recent work \citep[e. g.,][]{torres}.  The steady-state CR
diffusion-loss equation \citep{longair} is now integrated over energy,
where the original work solved a time integration, requiring a
slightly different Green's function.  For electrons, as before, energy
losses due to ionization, non-thermal bremsstrahlung, synchrotron
radiation, IC scattering, and escape are accounted for
\citep[][appendix]{n253.gammas, torres}.  The full Klein-Nishina cross
section is now included \citep[][p. 88]{schlick} for IC scattering 
since relativistic effects become important at the high energies to
which $Fermi$ and ground-based arrays are now sensitive.
Bremsstrahlung losses are now included in the continuous CR energy
loss rate rather than approximating it with an energy-independent CR
loss timescale.  Also the photon energy density for IC is now
constrained using the starburst templates of \citet{sbgrid} fit to the
IR spectra.  Finally, positrons (see below) are now included.

The CR protons are affected by ionization and escape losses, plus pion
production above a total energy of 1.22 GeV.  The cross section
parameterizations of \citet{kamae} are now employed for all pion 
species ($\pi^{\pm,0}$).  Secondary CR electrons are created via
knock-on or Coulomb interactions, and the decay of negatively charged
pions.  Secondary positrons are created through the decay of positive
pions.  The distributions of these secondary leptons from charged
pions are also determined using the parameterizations of
\citet{kamae}.

Power law primary CR electron and proton injection spectra of the form
$Q(E) = K E^{-s}$, in particles \percc\ \pers\ \pmev, are assumed,
where $E$ is the particle total energy.  The previous model used a
power law in kinetic energy, similar to other studies \citep{torres},
which has a conveniently simple overall normalization.  The total
energy is chosen in this work not only to be consistent with other
theoretical studies \citep*{lacki,bell}, but as a deliberate contrast
from the previous work to test the effect of a different injection
spectrum.  This choice mostly affects the proton spectrum at low
energies, and thus the pion spectra, which, via secondary production,
could alter both the radio and gamma-ray emission, particularly near
the pion bump.

This is strictly a one-zone model only simulating the expected nuclear
emission from starburst galaxies.  No diffusion to, or emission from,
the extended disk is considered.  There are a variety or reasons for
this choice besides basic simplicity.  Given that these sources are so
near the flux and angular resolution detection thresholds already, the
much lower emissivities from the disk are insignificant to $Fermi$.
We originally modeled the disk of NGC 253 \citep{n253.gammas} and
indeed found its gamma-ray emissivity to be negligible compared to the
central starburst, in the same way as its radio emission is much
reduced \citep{ulvestad}.  Recently \citet{torres.3d} estimated more
realistically the steep decline (by an order of magnitude) of the
CR proton density outside the inner starburst in NGC 253.  The diffuse
gamma-ray and radio emission from the Milky Way Galactic center also
outshine the disk \citep{crocker}.  Our nearest neighbor,the LMC, one
of the very few extragalactic sources with resolved gamma-ray
emission, similarly only shows emission from its most massive site of
star formation \citep{butt}.

Using a power law injection spectrum in total energy also requires a
new normalization.  As usual, the normalization $K$ is determined by
the supernova rate, $\Psi$, starburst volume, $V$, and the efficiency
of power transfer between the supernovae and CRs, $\eta$, such that
\begin{equation}
\int^\infty_{m_pc^2} Q(E) E_{kin}\ dE = \eta P \Psi/V\ \ .
\label{eq.normz}
\end{equation}

\noindent
The supernova energy, $P = 10^{51}$ ergs.  A minimum CR kinetic
energy, $E_{kin} \gsim\, 2 m_p v_s^2 \sim$ a few MeV, is required for
acceleration in a shock front given shock velocities of $10^4$\kms
\citep{bell}, and is much smaller than the proton rest mass energy.
The relative normalization between the proton and electron CR
distributions at high energies then depends on the injection spectral
index $s$ \citep[][p. 472]{bell,lacki,persic,schlick}, and as such is
no longer an independent free parameter:
\begin{equation}
N_p/N_e = (m_p/m_e)^{(s-1)/2}\ \ .
\label{eq.NpNe}
\end{equation}

\noindent
Typical values of $s=2.0$--2.4 yield proton-to-electron ratios of
50--200, which are consistent with Milky Way observations and
effectively constrain the choice of $s$.  Note that this is a
simplification which differs from the treatment of \citet{lacki}
\citep[see also discussion in][]{persic} in order to reduce the number
of free parameters.  This normalization mostly impacts the secondary
contribution to the CR electron+positron flux, and introduces
variations of at most factors of $\sim 2$ from other treatments.

A convection loss timescale $\tau_c = 1$ Myr is assumed, which is
consistent with estimates of starburst winds and CR transport time (or
length) scales \citep{heesen, murphy}.  So the total energy-dependent
escape timescale is the combined diffusion and convection terms, or
\begin{equation}
\tau_{esc}^{-1}(E) = \tau_0^{-1} \beta E_{_{\!\rm GeV}}^{0.5} + 
                     \tau_c^{-1}\ \ .
\label{eq.tesc}
\end{equation}

\noindent
In sum, the model variables are $s$, the volume-averaged gas (H$_2 +$
H) density $n$, magnetic field strength $B$, diffusion time scale
$\tau_0$, and photon field energy density $U_{ph}$ (from the
integrated starburst SED template).

\subsection{Model Results: Steady-State CR Distributions}

\begin{figure}
\plotone{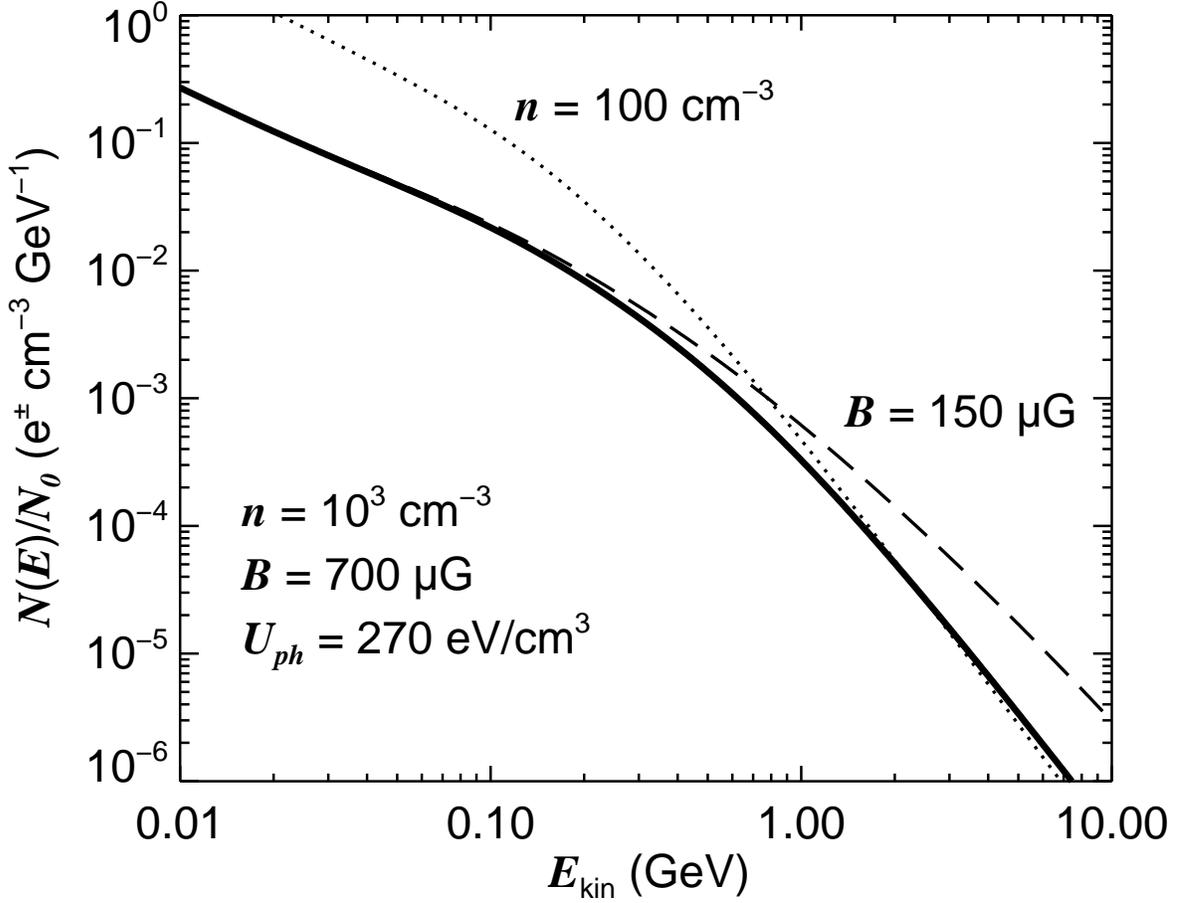}
\caption{Normalized total (secondary and primary) electron+positron
steady-state distributions.  The thick line shows model results for a
nominal photon field and relatively high gas densities and magnetic
field strengths.  The dashed line shows the effect of lowering 
the magnetic field, and the dotted line shows the effect of a low
density.  Note in the high density model the slightly more pronounced
bump around 0.1 GeV from enhanced high energy secondaries.
For all models here, $s = 2.2$ and $\tau_0 = 10$ Myr.
\label{fig.emodel}}
\end{figure}

Figure~\ref{fig.emodel} shows the results from a few models broadly
indicating the effects of varying density and magnetic field
strength on the steady-state CR electron+positron distributions.
Higher densities increase the ionization losses at low energies,
leading to an overall flatter, or more curved, spectrum.  High
magnetic fields increase synchrotron radiation losses, subsequently
decreasing the CR electron distribution, particularly at high
energies.  As a result, high energy emission from inverse Compton (IC)
scattering and non-thermal bremsstrahlung are reduced.  Losses from IC
scattering are less important; the overall results are quite
insensitive to changes in $U_{ph}$ or the FIR spectrum template.
Increased losses due to a shorter convection timescale ($\tau_c$) are
also relatively small, primarily lowering and slightly steepening the
CR proton spectrum around TeV energies.  Longer diffusion timescales
\citep[higher $\tau_0$, see][]{torres.snr} result in slightly harder
and higher CR spectra above 10 GeV or so.  This effect is also
comparatively small, reaching factors of just up to a few above a TeV,
and negligibly changing the radio spectrum.

Altering the injection slope $s$ affects the steady-state spectral
slopes accordingly, but can also substantially change the contribution
of secondary particles because of its effect on the proton-to-electron
ratio.  For example, for flatter slopes, secondary particles
contribute more to the lepton distribution (Eq.~\ref{eq.NpNe}),
causing a more pronounced bump around 0.1 GeV in the spectrum.  Note
that this secondary bump is very subtle compared to the primary
spectrum, so adjusting it by factors of a few changes the total lepton
spectrum very little.  Thus the results are not very sensitive to
whether a power law injection spectrum in total or kinetic energy is
chosen, nor to the specific treatment of $N_p/N_e$.

The CR proton spectrum depends primarily on the density.  As with the
electrons, higher densities slightly flatten the proton spectrum at
low energies from ionization losses.  However, high energy losses
are also increased via pion production, so the overall effect is that
the proton distribution decreases, and slightly flattens, with
density.  The effect of using an injection spectrum that is a power
law in total, rather than kinetic, energy mostly appears in the
resultant pion distributions (and thus the secondary electron+positron
distributions).  The pion spectrum is marginally lowered (by factors
of $\sim 2$) at the 0.1 GeV bump, but is unaffected at high energies.
This small difference seems to have a negligible effect on the
resultant gamma-ray and radio emission spectra as the secondary
contribution is slight, and the high energy pion distributions are
unaltered.

\subsection{Model Results: Gamma-ray and Radio Spectra}

Radio and gamma-ray spectra were generated from the steady-state CR
distributions.  Gamma-rays are created by IC scattering and
non-thermal bremsstralung from primary and secondary CR
electrons+positrons, plus the decay of neutral pions from CR protons
interacting with ambient gas.  Radio spectra are generated by
synchrotron radiation from these CR leptons, thermal free-free
absorption and emission, plus warm dust emission at millimeter
wavelengths.  We neglect TeV absorption and pair production for now as
their expected impact is relatively small and difficult to discern
with current sensitivities \citep{torres}.

\begin{figure}
\plotone{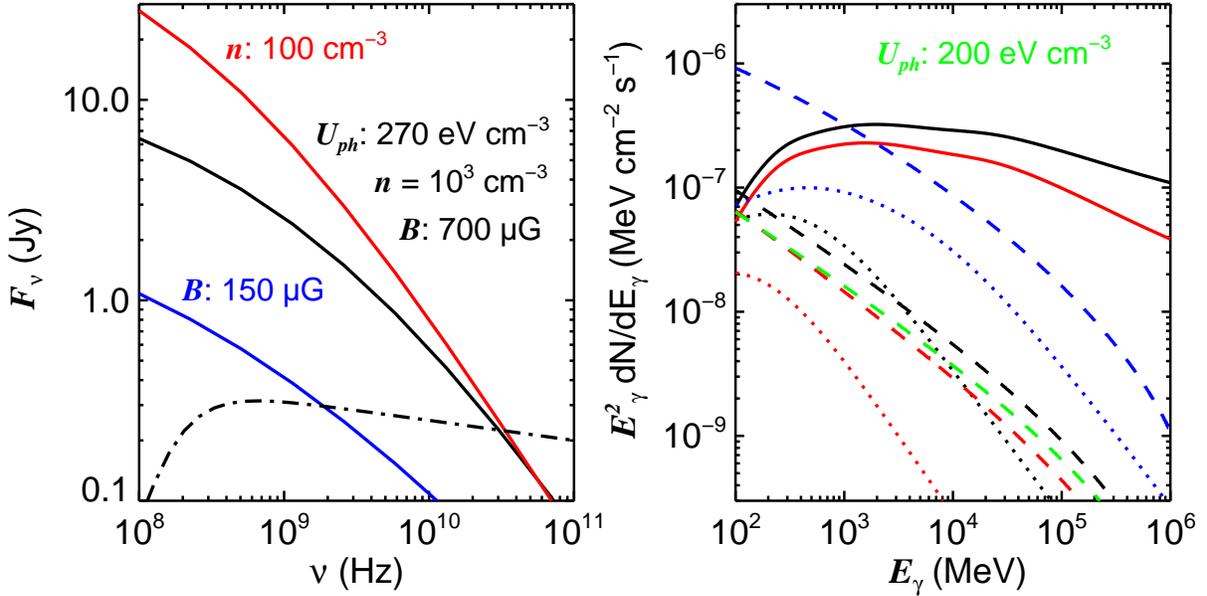}
\caption{Sample model results.  (Left) Expected radio synchrotron
emission for nominal values of density, magnetic field, and photon
density (black).  Here $s = 2.2$, $\tau = 10$ Myr, and $\eta = 0.02$.
The blue line is for a lower magnetic field, and the red line is for a
lower density.  The dot-dash line indicates a typical thermal
bremsstrahlung level and the onset of free-free absorption at low 
frequencies (here $\tau_{1 \rm GHz}$ = 0.04).  The thermal absorption
has not been applied to the synchrotron spectra shown here.  (Right)
Expected gamma-ray emission due to neutral pion decay (solid lines),
non-thermal bremsstrahlung (dotted lines), and IC scattering (dashed
lines).  Colors are as before.  The green line indicates a lower
photon density which is indistinguishable from the nominal case except
for IC scattering.
\label{fig.models}}
\end{figure}

The gamma-ray spectrum of a star forming galaxy above a GeV is
dominated by pion decay from CR protons interacting with the
interstellar medium (Figure~\ref{fig.models}).  At lower gamma-ray
energies, bremsstrahlung and/or IC scattering from CR electrons is
important.  The IC contribution is very sensitive to the high energy
tail of the CR lepton distribution, and contributes most when there is
a sufficient density of very high energy CRs, that is, when the magnetic
field -- and associated energy losses -- is low.  In fact, sensitive
MeV data would be very useful to constrain the TeV CR lepton
distributions in these galaxies.  Higher densities enhance not only
pion decay emission, but also non-thermal bremsstrahlung, since the
secondary CR electron and positron distributions increase with density
given that they too arise from CR protons (via charged pion decay).
Therefore, the gamma-ray spectrum normalization is most sensitive to
density, with some dependence on magnetic field strength below a GeV.

The radio spectrum normalization is very sensitive to the magnetic
field.  Higher gas densities tend to reduce the radio signal somewhat
by flattening the CR electron distribution.  The flattening, or
curvature, in the radio spectrum \citep{wb,thompson} is a result of
ionization reactions affecting mostly lower energy CR electrons.
Higher densities also enhance the bump feature in the secondary
distributions, adding to the curvature.  To a much lesser extent, the
far-infrared (FIR) emission from dust also affects the radio and
gamma-ray spectra by providing a deep reservoir of seed photons for IC
scattering.  So a high FIR luminosity increases CR leptonic losses,
consequently diminishing slightly the non-thermal bremsstrahlung and
synchrotron emission, while enhancing the IC gamma-ray flux.  The
FIR-radio correlation hints at the interrelatedness of all of these
processes in star-forming galaxies.  Neither the radio nor gamma-ray
emission is very sensitive to the diffusion timescale.  Together the
radio and gamma-ray normalizations effectively constrain $n$ and $B$.

\subsection{Model Fits to NGC 253 and M82}

\begin{figure}
\plotone{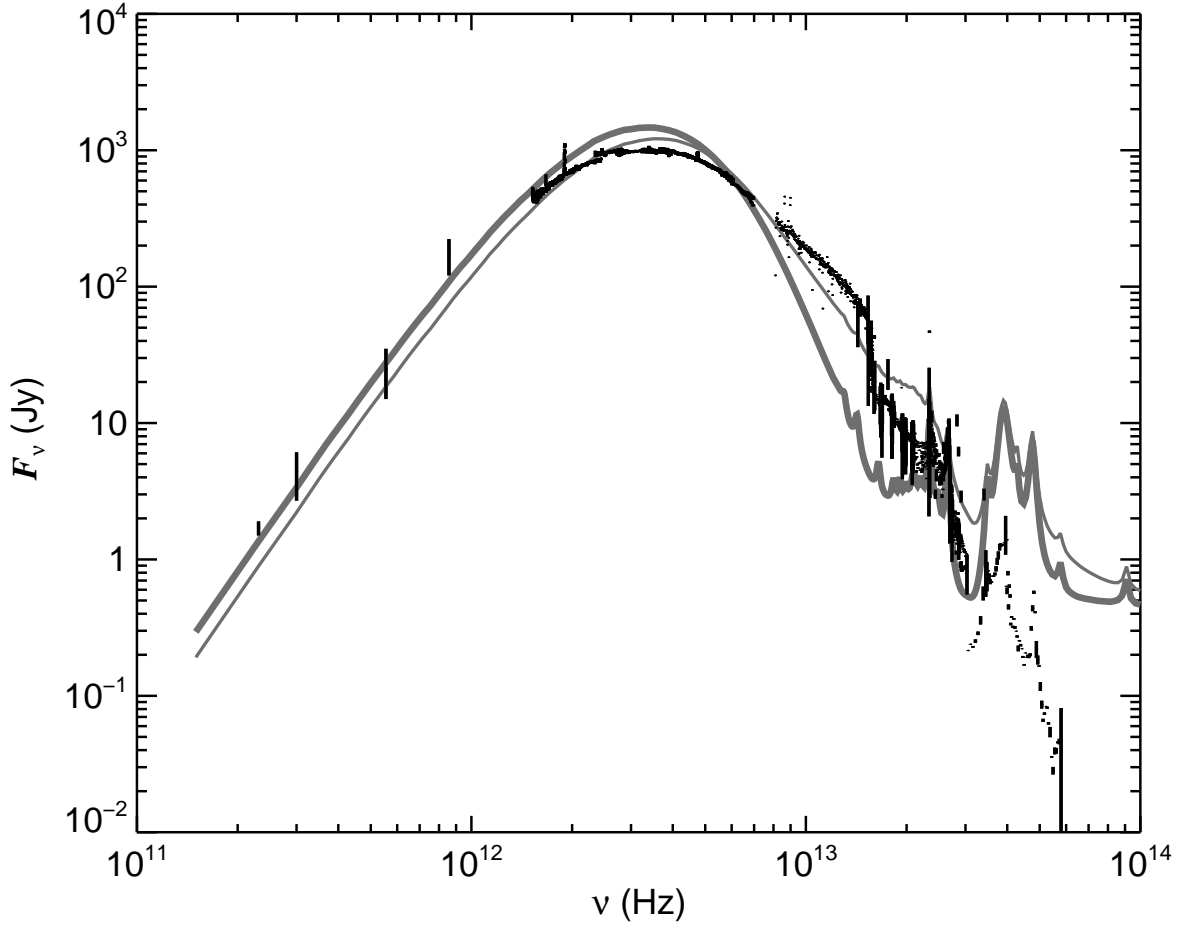}
\caption{IR spectrum of NGC 253.  The SED models shown are for a 350
pc radius starburst region, a ratio of OB luminosity to total
luminosity of 0.6, total IR luminosity of $10^{10.5} \lsun$, $A_V =
72$ mag, and a hot spot density of $10^4$ \percc\ (thin line).  The
thick line is for a similar model but with $L_{IR} = 10^{10.2} \lsun$
and a hot spot density of $10^2$ \percc.  References:
\citet{riekelowa, riekelowb, radovich, rieke350, hildebrand, chini,
krugel}.
\label{fig.ir253}}
\end{figure}

\begin{figure}
\plotone{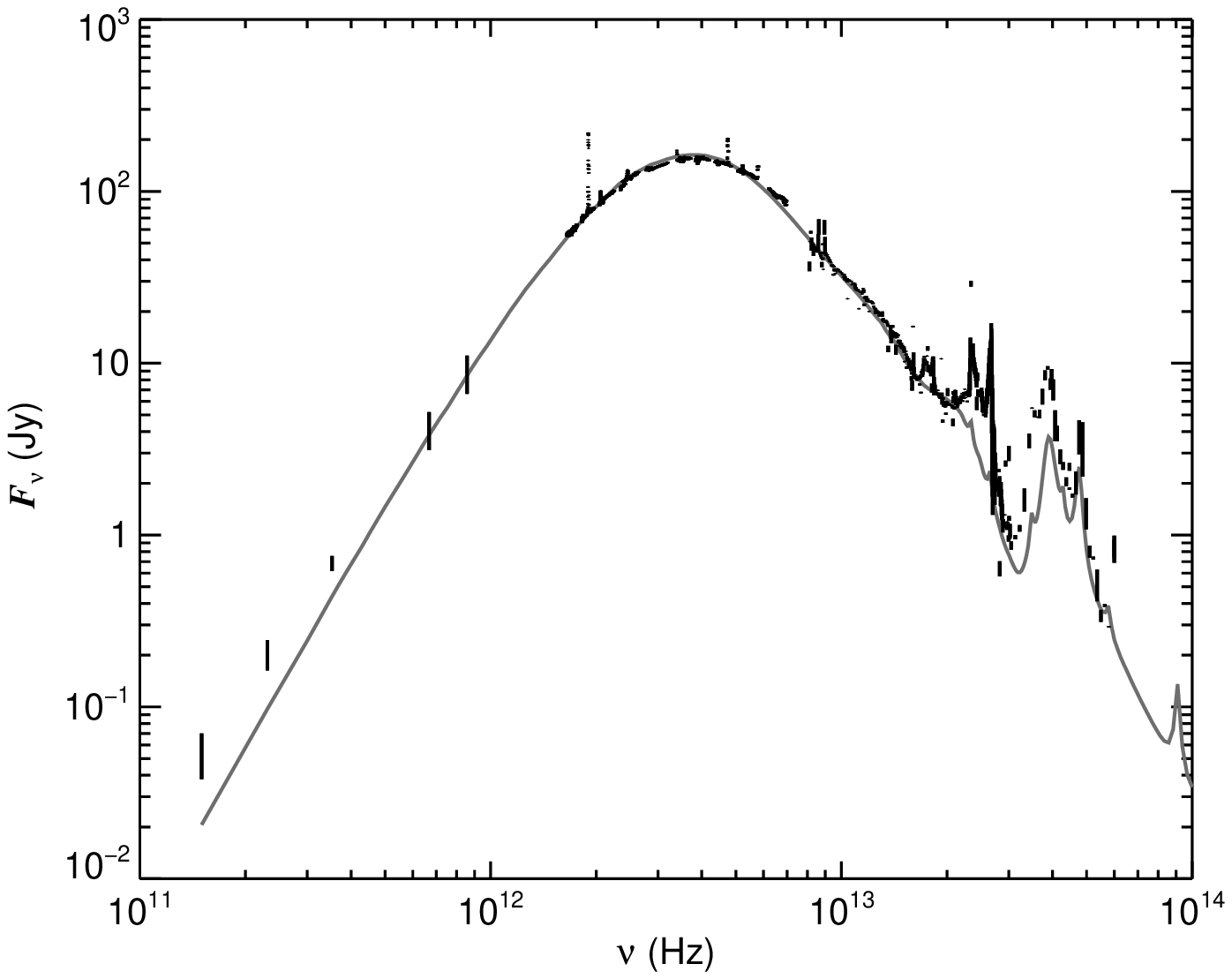}
\caption{IR spectrum of M82.  The SED model shown is for a 350 pc
radius starburst region, a ratio of OB luminosity to total luminosity
of 0.9, total IR luminosity of $10^{10.7} \lsun$, $A_V = 72$ mag, and
a hot spot density of $10^4$ \percc.  References:
\citet{riekelowa, kleinmannlow, isocam, riekelebofsky, rieke, alton}.
\label{fig.ir82}}
\end{figure}

The $Spitzer$ IR spectra of NGC 253 and M82 are shown in
Figures~\ref{fig.ir253} and \ref{fig.ir82} with some closely matching
starburst SEDs from \citet{sbgrid}.  The photon densities from these
models are 171 and 274 eV \percc\ for NGC 253 and M82, respectively.
The assumed supernova rates are 0.1 and 0.08 yr$^{-1}$ for M82 and
NGC 253, respectively, and their assumed distances are 3.9 and 3.5 Mpc.

The radio and gamma-ray spectra were simultaneously fit using hundreds
of model runs over a broad parameter space.  Given the possibility
that the magnetic fields in such strong starbursts may be quite high
\citep{thompson}, field strengths up to mG were included.  Despite
the small effect, we also investigated very short and long diffusion
timescales, from 0.1 to 30 Myr since recent gamma-ray studies of
supernova-cloud interactions indicate that long timescales reproduce
the observations best \citep{torres.snr}.  The best fitting models
were determined using the minimum $\chi^2$.  The radio data were 
selectively chosen to include only the central starburst disks (data
from within the inner arcmin, or central 300--500 pc where the
molecular and mm continuum emission are strongest), not the more
extended winds/halos. In starburst galaxies, this extended emission
tends to have a much steeper spectrum \citep{wb, heesen, elmouttie}
and is derived from a more evolved CR population rather than one
directly interacting with the central starburst and contributing to
the pion decay gamma-rays. The volumes for these galaxies were assumed
to be disks of 150 pc radius and 70 pc thickness.  This volume may
underestimate their full molecular extent, especially in M82, but this
distinction is not so critical here; fluxes are generated from the
modeled emissivities by multiplying by the volume \citep{torres}, thus
canceling with the $1/V$ term in the normalization
(Equation~\ref{eq.normz}).

\begin{figure}
\plotone{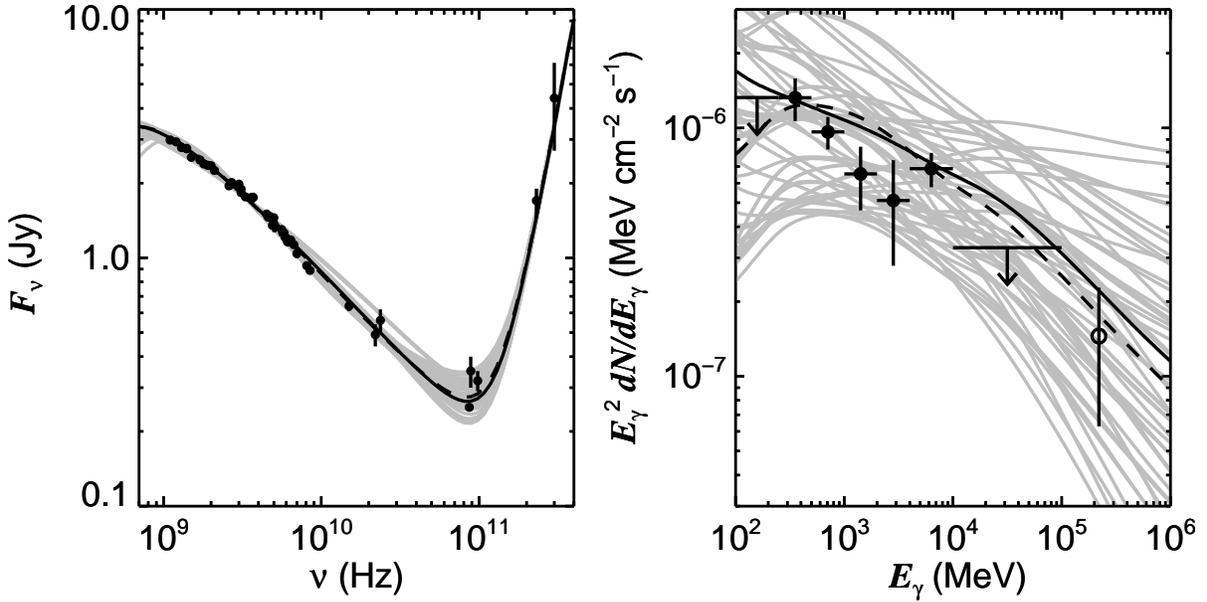}
\caption{Radio (left) and gamma-ray (right) spectra from the central
few hundred pc of NGC 253.  Model solutions within 3$\sigma$ of the
minimum $\chi^2$ are shown in grey with the minimum ($\chi^2 = 2.2$)
in black.  The best solution given the priors $\tau_0=10$ Myr and
$n > 300$ \percc\ is indicated with the dashed line ($B = 350
\mu$G, $n = 10^3$ \percc, $s=-2.4$, $U_{ph}=200$ \percc, $\eta =
0.05$, $\chi^2=2.6$.).  Radio references: (NGC 253 core) \citet{wb,
carilli96, carilli92, ricci, takano, n253.ihcop, carlstrom, n253.cs}.
\label{fig.n253}}
\end{figure}

\begin{figure}
\plotone{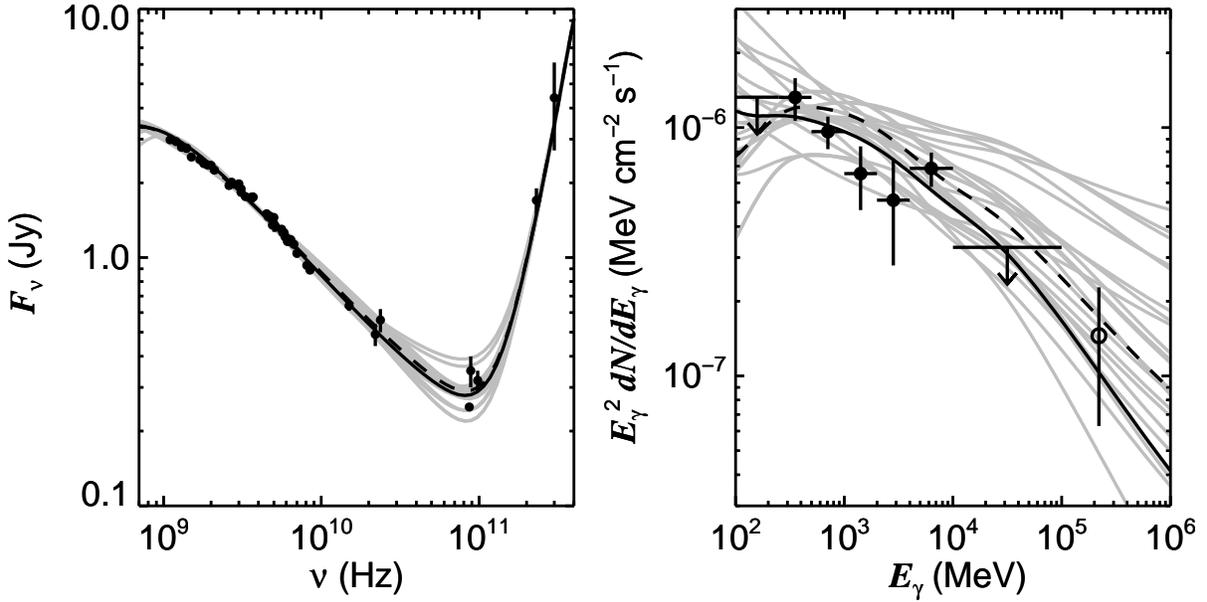}
\caption{Same as Figure~\ref{fig.n253}, but after increasing the
weighting of the gamma-ray data by the ratio of the numbers of radio
and gamma-ray points (minimum $\chi^2 = 4.9$).  The best solution for
the priors $\tau_0=10$ Myr and $n > 300$ \percc\ is indicated with
the dashed line ($B = 350 \mu$G, $n = 10^3$ \percc, $s=-2.4$,
$U_{ph}=200$ \percc, $\eta = 0.05$, $\chi^2=5.3$.).
\label{fig.n253w}}
\end{figure}

\begin{figure}
\plotone{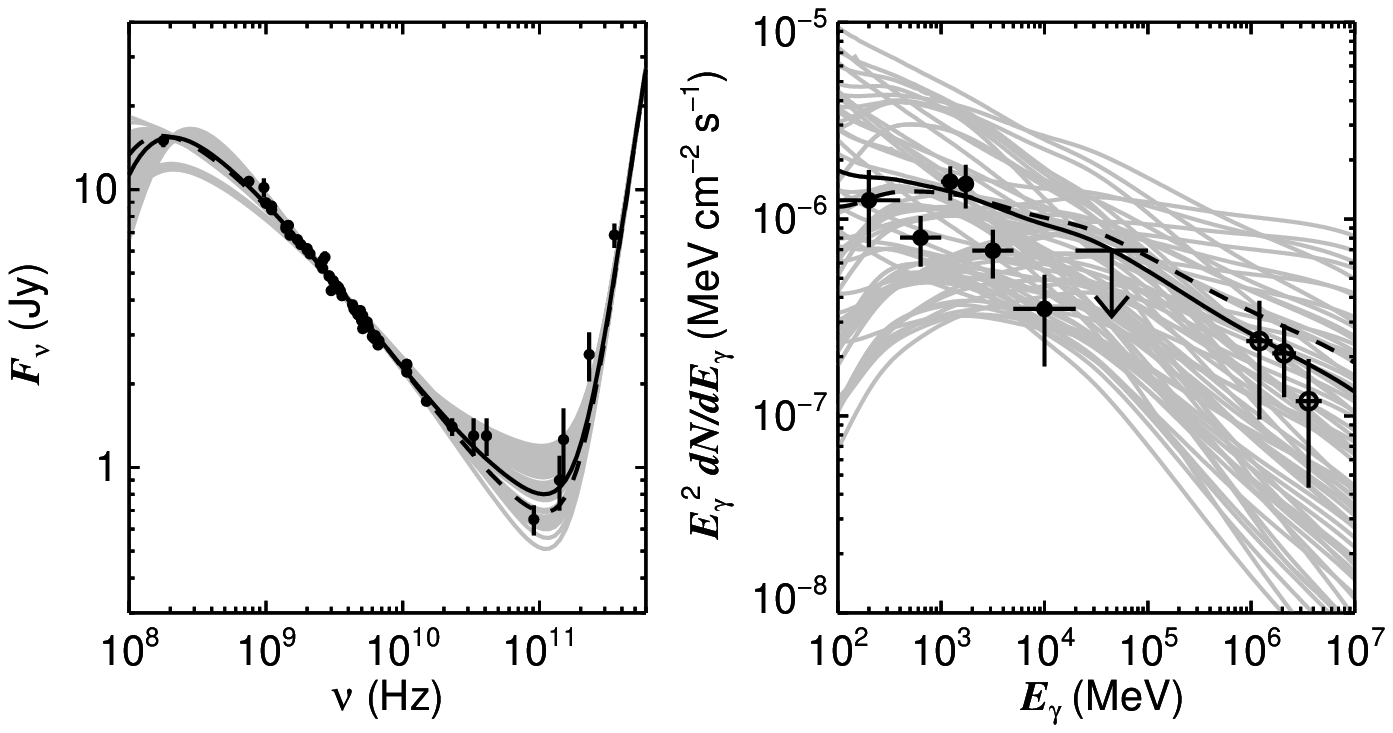}
\caption{Same as Figure~\ref{fig.n253}, but for M82 (minimum $\chi^2 =
3.5$).  The best solution for the priors $\tau_0=10$ Myr and $n >
300$ \percc\ is indicated with the dashed line ($B = 450 \mu$G, $n
= 600$ \percc, $s=-2.2$, $U_{ph}=270$ \percc, $\eta = 0.07$,
$\chi^2=3.7$.).  Radio references: \citet{wb, hughes, wmap,
laingpeacock, kellermann, pauliny, kuhr, condon}.
\label{fig.m82}}
\end{figure}

The best fit models are shown with the data in
Figures~\ref{fig.n253}--\ref{fig.m82}.  There is a large mismatch in
both the noise and number of data points between the radio and
gamma-rays.  Thus the scatter in the gamma-ray models is
understandable since the normalization is dictated mostly by the radio
data.  To stress the physical implications of the new $Fermi$ data,
fits were done artificially increasing the weighting of the gamma-ray
data by the ratio of the numbers of radio and gamma-ray points.  Very
similar solutions result, though over a narrower range of parameter
space.

\citet{lacki11} compared the first $Fermi$ catalog detections and TeV
data to a collection of published model predictions and concluded that
they were for the most part broadly consistent with each other (many
of these models were constrained merely by non-detections at GeV
energies and so considerably overestimated the observations).  The
scatter in the models in their Fig.~1 mirrors the 3-$\sigma$ envelope
of fits in Figs.~\ref{fig.n253}--\ref{fig.m82}.

%\bibitem[Rephaeli et al.(2010)]{rephaeli} Rephaeli, Y., Arieli, Y.,
%\& Persic, M.\ 2010, \mnras, 401, 473
%\bibitem[de Cea del Pozo et al.(2009)]{de Cea del Pozo} de Cea del
%Pozo, E., Torres, D.~F., \& Rodriguez Marrero, A.~Y.\ 2009, \apj,
%698, 1054 
%\bibitem[Domingo-Santamar{\'{\i}}a \&
%Torres(2005)]{domingosantamaria} Domingo-Santamar{\'{\i}}a, E., \&
%Torres, D.~F.\ 2005, \aap, 444, 403 

Normalizing the radio and gamma-ray models to match the spectra yields
power transfer efficiencies $\eta \approx 6\%$ for both galaxies.  All
fits within 3-$\sigma$ of the minimum $\chi^2$ result in efficiencies
well below 20\%, which is consistent with theoretical predictions
\citep{eta}.

\begin{figure}
\plotone{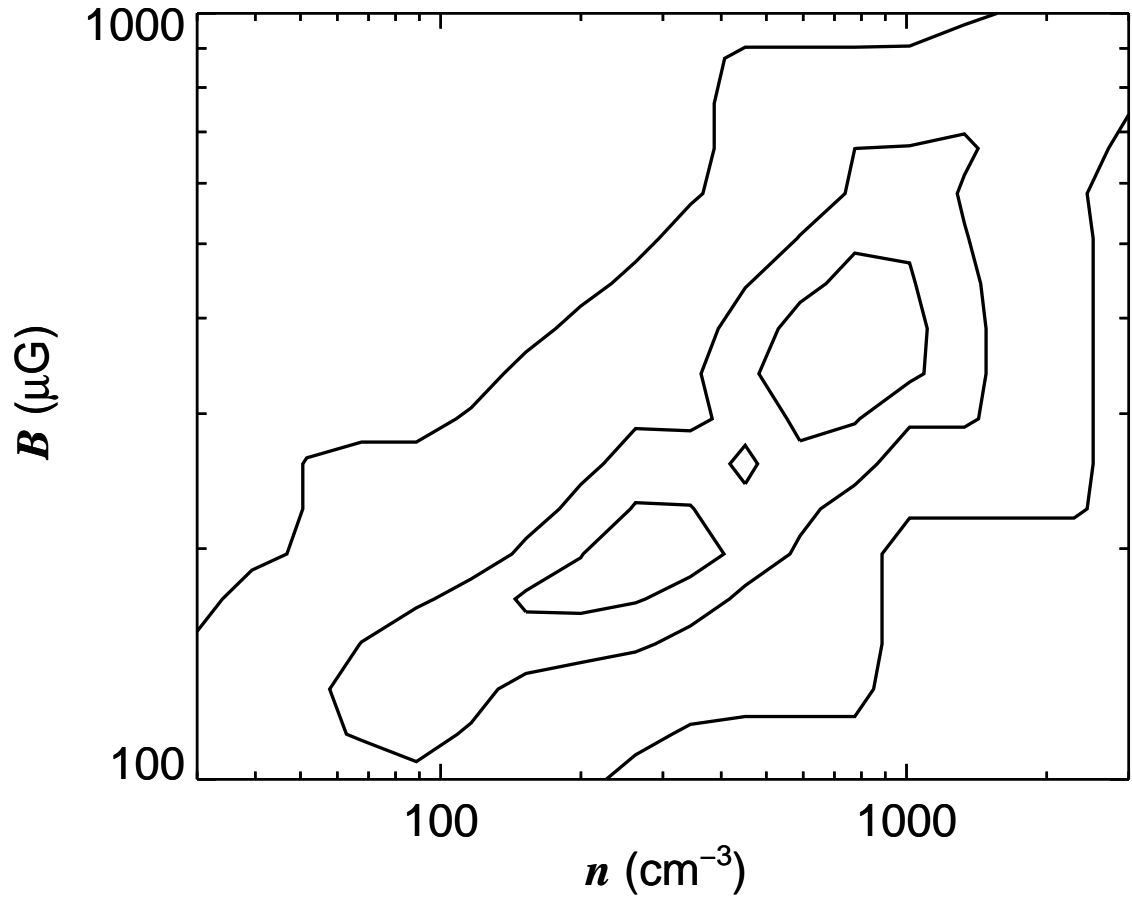}
\caption{Density and magnetic field solutions for the best fitting
models for NGC 253.  Here $s = 2.2$, $U_{ph} = 200$ eV \percc, and
$\tau_0 = 10$ Myr.  Contours indicate the 67\%, 90\%, and 99\%
confidence levels around the minimum $\chi^2=2.5$.
\label{fig.dm253}}
\end{figure}

\begin{figure}
\plotone{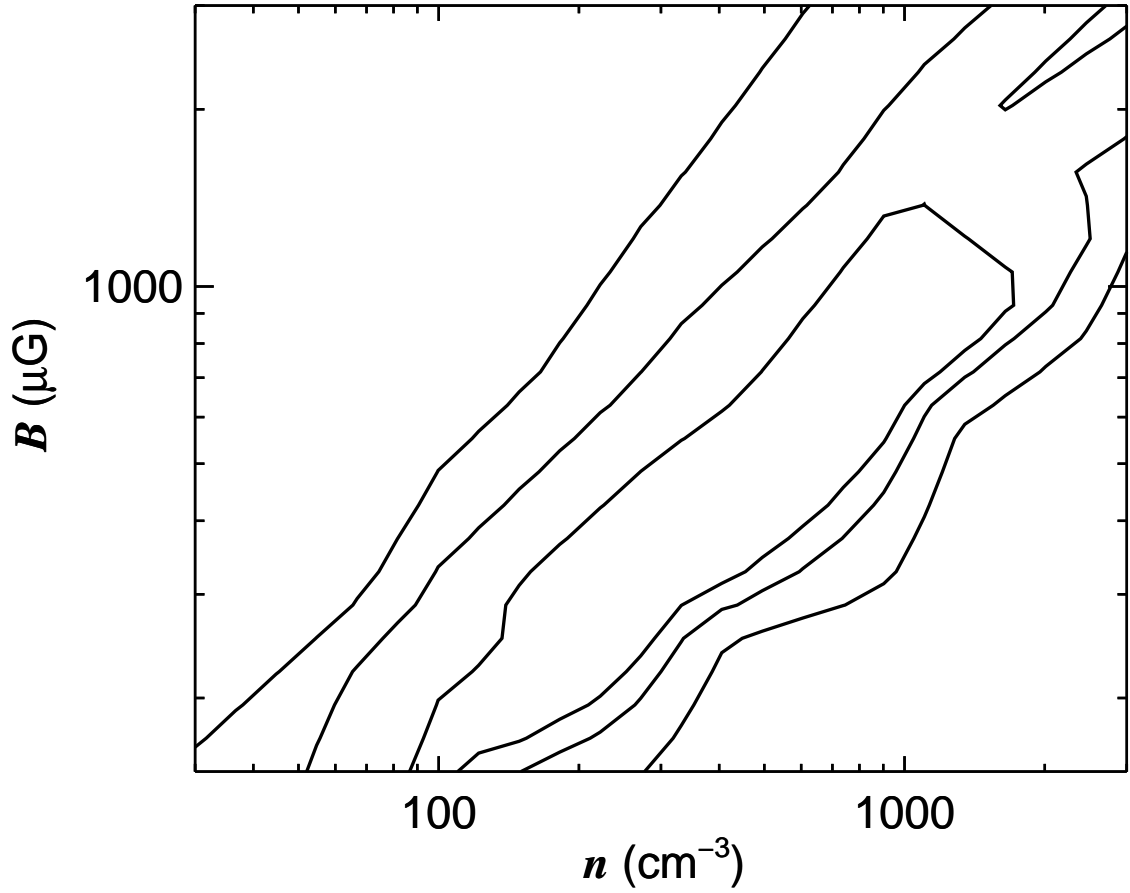}
\caption{Model solutions for M82 as in Fig~\ref{fig.dm253}.  Here $s =
2.2$, $U_{ph} = 270$ eV \percc, and $\tau_0 = 10$ Myr.
\label{fig.dm82}}
\end{figure}

Solutions with low $\chi^2$ values (Figures~\ref{fig.dm253} and
\ref{fig.dm82}) do include low magnetic fields around the minimum
energy value, $B \sim 150 \mu$G, but with very low gas densities of
$\sim 100$ \percc.  Based on studies of dust and molecular gas in
these galaxies, volume-averaged densities of at least several hundred
H$_2+$H \percc\ are expected \citep{n253.co, sakamoto, m82.co}.
Models with higher densities imply higher magnetic fields of several
hundred $\mu$G, approaching the calorimeter model expectations.  Valid
solutions also result for short confinement timescales.
\citet{torres.snr} found that a diffusion coefficient of $D = 10^{26}$
cm$^2$ \pers\ is strongly constrained by $Fermi$ and TeV observations
of supernova-cloud interactions.  Given $\tau_0 \approx R^2/2D$,
and setting $R$ to the starburst radius of 150 pc, yields $\tau_0 = 8$
Myr.  We restrict the following discussion just to models where $n >
300$ \percc\ and $\tau_0=10$ Myr.

\section{Discussion}

\begin{figure}
\plotone{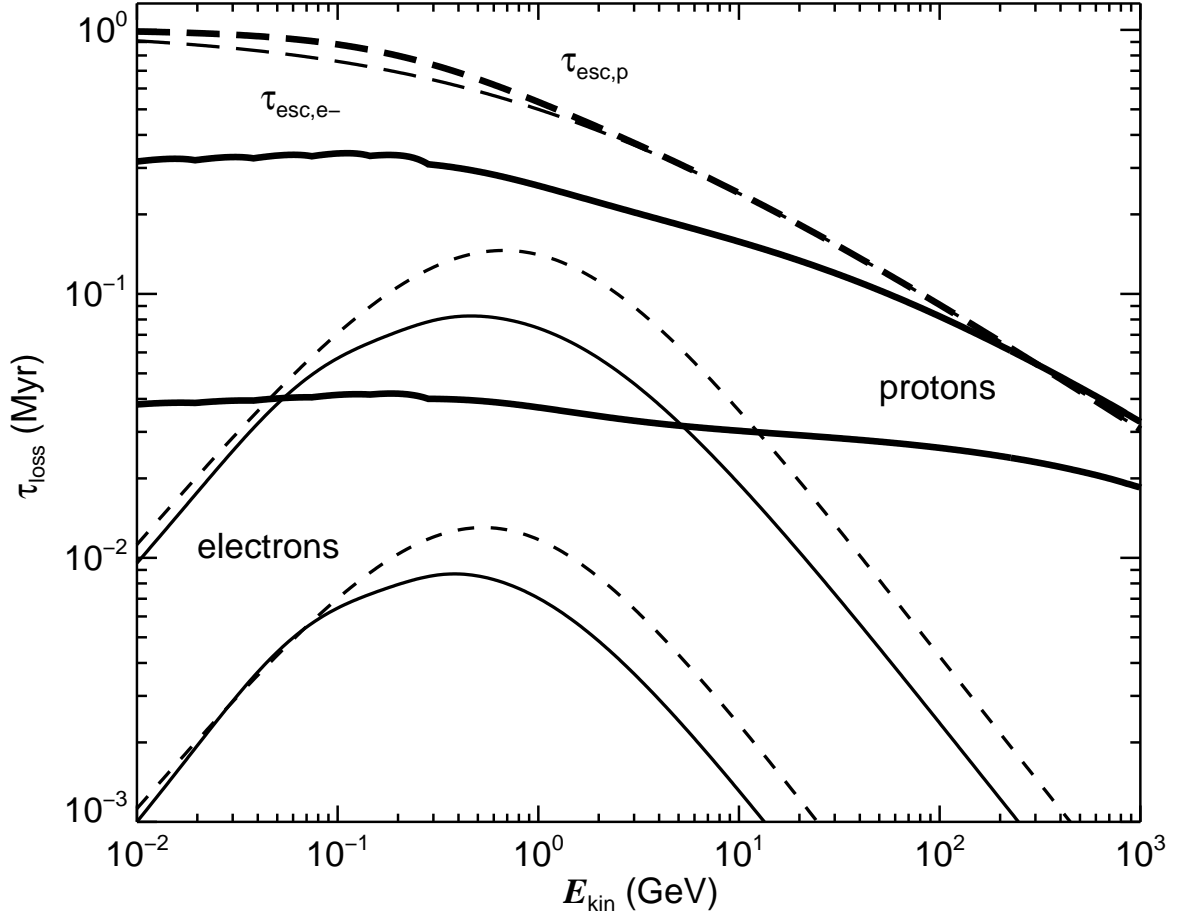}
\caption{CR electron and proton loss and escape timescales.  Electron
energy loss timescale, $E/b(E)$ (short dashed lines); electron
steady-state lifetime, $N(E)/Q(E)$ (solid lines); electron escape
timescale, Eq.~\ref{eq.tesc} (long dashed line).  Here $s = 2.2$,
$U_{ph} = 270$ eV \percc, and $\tau_0 = 1$ Myr.  Upper curves are for
slow loss rates, $n = 100 \percc$ and $B = 150 \mu$G; lower curves are
for fast loss rates, $n = 10^3$ \percc\ and $B = 700 \mu$G.  Proton
loss ($N/Q$ only) and escape timescales are denoted with thick lines.
Again, the upper and lower curves are for $n = 100$ and $10^3$ \percc,
respectively.
\label{fig.losses}}
\end{figure}

These results generally favor the high magnetic field calorimeter
model for starburst CR populations \citep{thompson}.  The implication
is that the CR cooling rate, primarily via synchrotron radiation and
ionization losses, is faster that the escape rate.
Figure~\ref{fig.losses} shows the loss timescales for electrons using
models consistent with the radio and gamma-ray data for NGC 253.  Here
the CR lifetime is simply approximated as either the energy loss
timescale $E/b(E)$, where $b(E)$ is the CR energy loss rate
\citep{torres,n253.gammas}, or the steady-state lifetime,
$N(E)/Q(E)$.  Even for lower densities and magnetic fields, the
electron loss timescales are shorter than the escape timescale.  Even
protons appear mostly calorimetric, especially for low energies and
high densities.  So it would seem that the condition that $\tau_{loss} 
\ll \tau_{esc}$ is satisfied for electrons, particularly for
higher values of $n$ and $B$ and given that diffusion timescales are
likely longer than 1 Myr (recall that $N$ depends weakly on $\tau_0$).
Therefore the minimum energy estimate for magnetic field strength may
indeed underestimate the true value in starburst galaxies.

Another argument in favor of the high field (high density) solutions
comes from a rough examination of the transport and interactions of
CRs during their lifetimes within the starburst region.  CRs encounter
a column of primarily molecular gas which has been measured from CO
studies to have beam-averaged values of up to $10^{22}$ H$_2$ \cmtwo\
\citep{n253.co,umigs,m82.co}, where typical single-dish beam sizes 
encompass the entire central starburst region (300--600pc).  One may
approximate this column density as $n\Delta l$, where $\Delta l$ is
akin to the mean free path of the CR.  This path length is then of
order 100 pc for low average gas densities (30 \percc) and around a
few pc for high values ($10^3$ \percc).  The former result implies
that the typical CR leaves the starburst region within its lifetime,
rather than losing the majority of its energy within it, contradicting
the steady-state assumption and calorimetry.  But it is hard to
understand independent studies of very warm cloud core temperatures in
starbursts if CRs interact in such a limited way with molecular cloud
cores (explored below).  The high density (and high magnetic field)
solutions imply that CRs lose the majority of their energy on spatial
scales of order the sizes of dense molecular cloud clumps or cores.
This result is much more consistent with model expectations and
observations from photodissociation and/or cosmic ray dominated
regions \citep{matsushita, papa}, the calorimeter model,
the high measured cloud temperatures \citep{panuzzo}, and the size
scales of Galactic molecular clumps at such densities \citep{GRS}.
Confirming this expectation are the measured volume densities and peak
column densities of individual, $\sim 30$ pc scale cloud complexes in
NGC 253 and M82 ($10^{22-24}$ H$_2$ \cmtwo\ and $10^{3-4}$ \percc)
\citep{sakamoto,m82.co}.  Comparable magnetic field strengths, and
volume and column densities, have been inferred across the full $\sim
40$ pc extent of the massive star-forming cloud Sgr B2 in the Galactic
center \citep{crutcher.sgrb}.

Further justification for the high density and magnetic field
solutions may be derived from the expected ionization rates from
penetrating CRs in starburst molecular clouds.  Ionization rates,
$\zeta = \zeta_e + \zeta_p$, are estimated using the best fit models
for CR electron+positron and proton distributions, and the cross
sections and procedures from \citet*{padovani}.  Only the H$_2$ cross
sections are used in this analysis since they are the most
significant.  The steady-state CR fluxes are extrapolated down to the
H$_2$ ionization energy of 15.6 eV for the integration.  Note that
this extrapolation overestimates the true ionization rate since other
absorption processes such as H$_2$ excitation and dissociation may
occur as gas column densities get large.  \citet{padovani} indicates
that at low energies ($< 1 $ MeV and $< 100$ MeV for electrons and
protons, respectively), the CR spectrum becomes rather flat, with a
plateau that declines with the encountered column density.  Deep
within clouds, the CR proton contribution dominates over CR electron 
ionization given the apparently more gradual decline of $\zeta_p$ with
column density.

\begin{figure}
\plotone{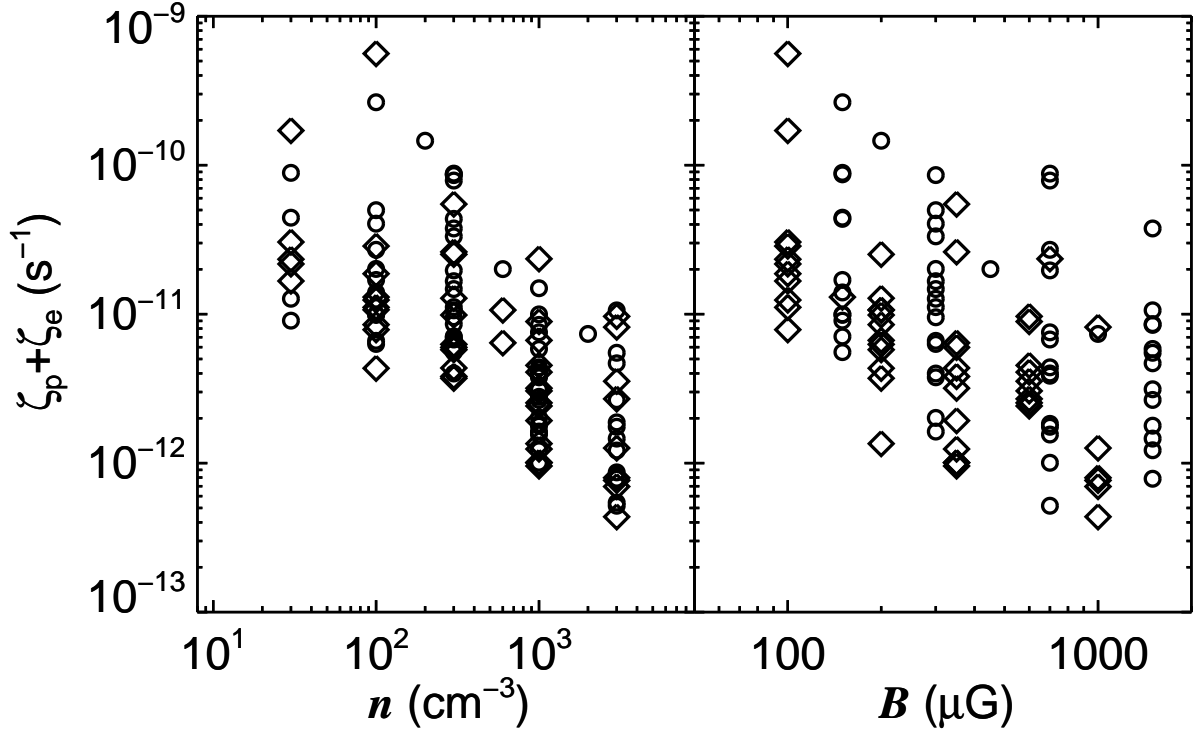}
\caption{Ionization rates for M82 (circles) and NGC 253 (diamonds).
Note that $\zeta$ here is expected to overestimate the actual
ionization rate given the unmodeled attenuation of the low energy CR
fluxes as they pass through molecular clouds.
\label{fig.zeta}}
\end{figure}

\begin{figure}
\plotone{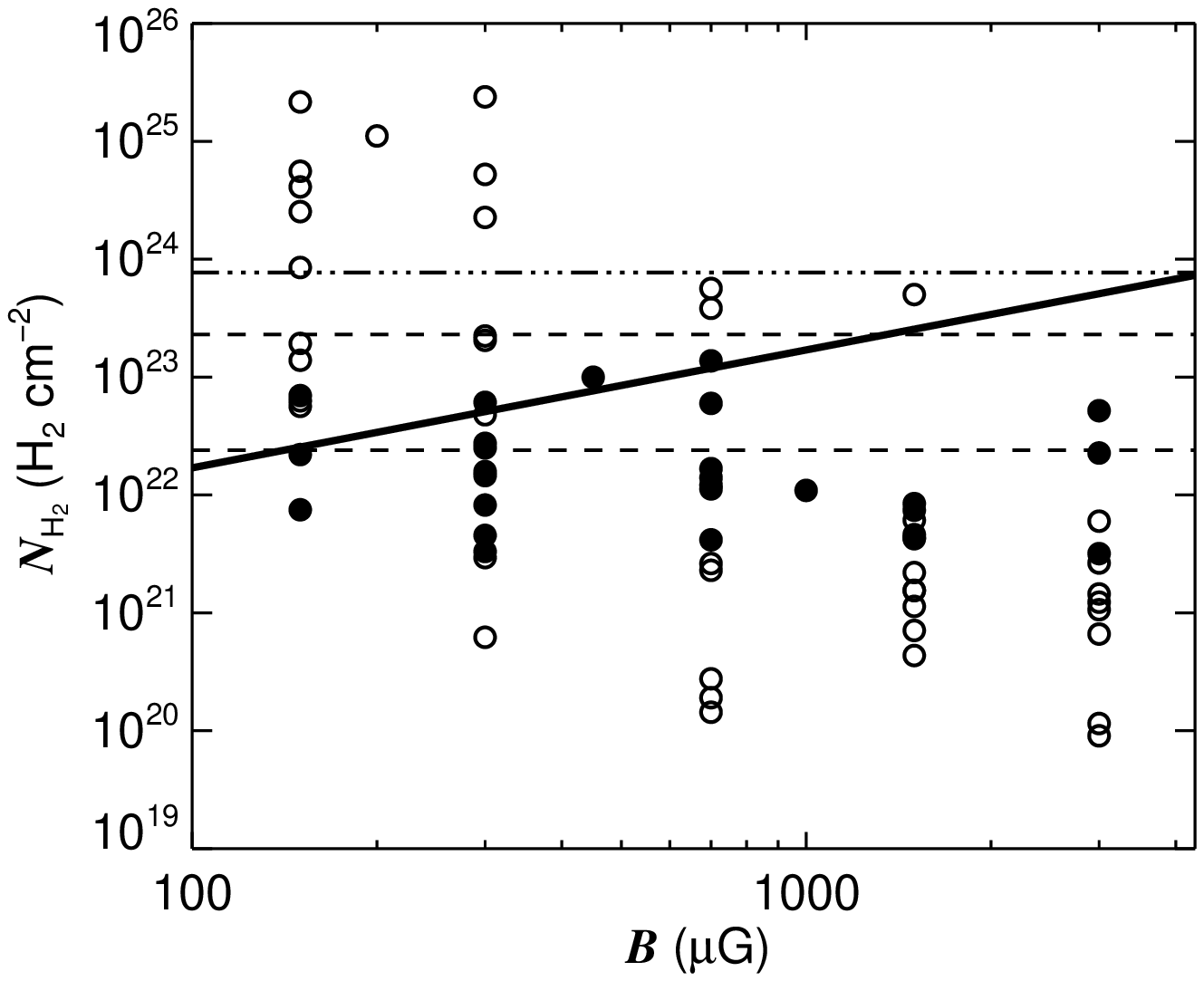}
\caption{Column densities needed to attenuate the ionization rates in
M82 to the fiducial value of 4\ee{-15} \pers. Dashed lines indicate
the range of column densities estimated from CO observations
\citep{m82.co}, and the dash-dot line is deduced from the central
dynamical mass \citep{panuzzo}.  The thick line indicates the relation
between column density and magnetic field in Galactic molecular clouds
\citep{crutcher}.  Filled symbols denote column densities resulting in
CR path lengths 1 pc $< \Delta l = N_\htwo/n <$ 200 pc.
\label{fig.zcol}}
\end{figure}

This analysis yields median values of $\zeta \approx 6\ee{-12}$ \pers\
for NGC 253 and M82 (Figure~\ref{fig.zeta}).  This CR ionization rate 
is a few orders of magnitude above the estimate in M82 based on CR
heating and the observed cloud temperatures, $\zeta \sim 4\ee{-15}$
\citep{suchkov}, and factors of 1--10 above the rate for ULIRGs
\citep{papa}.  Taking $\zeta$ from \citet{suchkov} as the
fiducial rate necessary to achieve the observed cloud kinetic
temperatures in M82 and NGC 253, and reducing the mean $\zeta$ from
this work by that magnitude requires a column density of $\sim
10^{22}$ H$_2$ \cmtwo\ \citep{padovani}, consistent with the observed
column density.  Further, Figure~\ref{fig.zeta} indicates trends
between $\zeta$ and density and magnetic field strength.  Solutions
for lower $n$ and $B$ require more attenuation of the low energy CR
flux to match the fiducial $\zeta$.  From \citet{padovani}, column
densities of up to $10^{24}$ H$_2$ \cmtwo\ or more are needed to
reduce $\zeta$ by the required factors of $10^4$
(Figure~\ref{fig.zcol}).  Such column densities exceed the gas mass of
their entire starburst regions, which further supports the higher
density solutions, and thus the high $B$ calorimeter model.  At very
high densities however, lower columns are needed to reach the fiducial
$\zeta$, and $\Delta l = N_\htwo/n$ can yield path lengths much less
than a pc, which is unreasonably small.  Therefore, the ionization
rate analysis disfavors the extremely high and low density (and
magnetic field) solutions.

Finally, theoretical and observed scaling relations between $B$, and
$N_\htwo$ and $n$, complete the argument against very low magnetic
fields in these starbursts.  While the radio and gamma-ray solutions
indicate a degeneracy between $n$ and $B$ (Figures~\ref{fig.dm253}
and \ref{fig.dm82}), which also roughly follows the expected scaling
relation $B \propto \sqrt{n}$, the magnetic field should also increase
proportionately with $N_\htwo$ \citep{crutcher}.  This scaling is
shown in Figure~\ref{fig.zcol}.  Again, magnetic fields of several
hundred $\mu$G are favored.

\section{Conclusions}

Self-consistent models of the steady-state CR distributions in
starburst galaxies match both the radio and gamma-ray spectra of NGC
253 and M82.  The models solve the CR diffusion-loss equation given a
power law spectrum of supernova-accelerated particles, accounting for
energy and particle losses due to ionization, synchrotron radiation,
IC scattering, bremsstrahlung, secondary particle generation, and
escape.  The radio normalization depends mostly on the magnetic field,
with some sensitivity to density in mostly the curvature of the
spectrum.  The gamma-ray spectrum, now observable with sufficient
sensitivity with $Fermi$ and TeV detections, is mostly normalized by
the density, with some dependence on the magnetic field below a GeV.
Therefore, with both radio and gamma-ray data, the densities and
magnetic fields in a starburst region may be independently and less
ambiguously constrained than from radio data alone.  In particular,
the question of calorimetry (high magnetic fields) in massive
starbursts may be addressed.

In NGC 253 and M82, these model results favor the high magnetic field
solutions anticipated by \citet{thompson}.  The higher densities also
implied are consistent with molecular gas studies.  Further, the
ionization rates, approximated from the steady-state CR distributions
with high magnetic fields and densities, also match expectations.
These results confirm that CR penetration and heating is an important
contributor to the warm temperatures observed in starburst galaxies.
These starburst nuclei may be populated to a significant filling
factor with giant molecular clouds resembling Sgr B2 in magnetic field
($\sim 500 \mu$G), column density ($10^{23}$ \htwo\ cm$^{-2}$), and
density ($10^3$ \percc).

\acknowledgments

The authors thank the $Fermi$ LAT team for their support, particularly
at local workshops and through the $Fermi$ Science Center website.  We
are especially grateful to T. Kamae for his guidance.  We thank
Patrick Lii for his early contributions to the starburst modeling
code, and the CUNY/AMNH REU program that sponsored him (NSF
AST-0851198).  We thank A. Marscher and A. Carrami\~nana for helpful
discussions.  This work was supported in part by the NASA New York
Space Grant Consortium based at Cornell University (\# NNX10AI94H).
This work was supported in part by grant \# 63388-00 41 from the
Professional Staff Congress of the City University of New York.  The
authors especially thank the referee for vital suggestions that
greatly strengthened this work.  This research has made use of the
NASA/IPAC Extragalactic Database (NED) which is operated by the Jet
Propulsion Laboratory, California Institute of Technology, under
contract with the National Aeronautics and Space Administration.  This
work is based in part on observations made with the Spitzer Space
Telescope, obtained from the NASA/ IPAC Infrared Science Archive, both
of which are operated by the Jet Propulsion Laboratory, California
Institute of Technology under a contract with the National Aeronautics
and Space Administration.

{\it Facilities:} \facility{Fermi (LAT)}

\clearpage

\clearpage

\begin{deluxetable}{ccccccc}
\tablewidth{0pt}
\tablecaption{Fermi LAT Differential Spectra of NGC 253 and M82
\label{tab.spec}}
\tablehead{
\multicolumn{3}{c}{M82} & & \multicolumn{3}{c}{NGC 253} \cr
\cline{1-3} \cline{5-7}
$E$ range & $E^2 dN/dE$ & TS && $E$ range & $E^2 dN/dE$ & TS \cr
(GeV) & ($10^{-6}$ MeV \cmtwo\ \pers) & && (GeV) & 
($10^{-6}$ MeV \cmtwo\ \pers) & 
}
\startdata
$0.1-0.4$    & $10.9\pm 4.0$\phn& 17 &&\phn$0.1-0.25$& $<9.47$        & 11 \cr
$0.4-1.0$    & $1.47\pm 0.34$ & 24 && $0.25-0.5$\phn& $2.60\pm 0.51$ & 25 \cr
$1.0-1.5$    & $0.52\pm 0.10$ & 53 && $0.5-1.0$     & $0.95\pm 0.14$ & 27 \cr
$1.5-2.0$    &\phn$0.26\pm 0.063$& 36 && $1.0-2.0$  & $0.36\pm 0.10$ & 20 \cr
$2.0-5.0$    &\phn$0.26\pm 0.058$& 47 && $2.0-4.0$ &\phn$0.17\pm 0.048$& 29 \cr
$5.0-20$\phd & $0.051\pm 0.024$&16 && $4.0-10$\phd &\phn$0.12\pm 0.019$& 25 \cr
\phn\phd$20-100$\phn & $<0.025$& 2.2&&\phn\phd$10-100$ & $<0.024$   & 1.6\cr
\enddata

\end{deluxetable}

\end{document}